\documentclass[structabstract]{aa}  

\usepackage{graphicx}
\usepackage{txfonts}
\usepackage{natbib}
\bibpunct{(}{)}{;}{a}{}{,} 

\begin{document}

   \title{Spatial distribution of stellar populations in the Magellanic Clouds: Implementation to Gaia}
   \titlerunning{Spatial distribution of stellar populations in the Magellanic Clouds}

   \author{M. K. Belcheva     \inst{1}   \and
           E. Livanou         \inst{1}   \and
           M. Kontizas        \inst{1}   \and
           G. B. Nikolov      \inst{1,2} \and
           E. Kontizas        \inst{3}
          }

   \institute{Department of Astrophysics Astronomy \& Mechanics, Faculty of Physics, University of Athens, GR-15783 Athens, Greece
              \and
              Department of Astronomy, Sofia University St.Kliment Ohridski, BG-1164 Sofia, Bulgaria
              \and
              IAA, National Observatory of Athens, PO Box 20048, GR-11810 Athens, Greece
              }

   \date{Received ; accepted }

   \abstract
    {}
    {The main goal of our project is to investigate the spatial distribution of different stellar populations in the Magellanic Clouds. The results from modelling the Magellanic Clouds can also be useful for simulations during the Gaia mission preparation.}
    {Isodensity contour maps have been used in order to trace the morphology of the different stellar populations and estimate the size of these structures. Moreover, star density maps are constructed through star counts and projected radial density profiles obtained. Fitting exponential disk and King law curves to the  spatial distribution allows us to derive the structural parameters that describe these profiles.}
    {The morphological structure and spatial distributions of various stellar components in the Magellanic Clouds (young and intermediate age stars, carbon stars), along with the overall spatial distribution in both Clouds, are provided.}
    {}
    
   \keywords{galaxies: Magellanic Clouds
            }

\maketitle

\section{Introduction}
The structure of the Magellanic Clouds has been a subject of study for many years. \citet{2001AJ....122.1807V} and \citet{2001AJ....122.1827V} determined the viewing angles of the LMC and constructed a near-IR star count map,  demonstrating that the LMC is intrinsically elongated. Many articles of the past decades \citep{1980A&A....87...92B,1996ApJ...461..742D,1992MNRAS.257..195G} also discuss the different area distributions of the various stellar populations in the MCs. \citet{1998A&A...338L..29M}, \citet{2000ApJ...534L..53Z}, and \citet{2001A&A...379..864M} show that, in both LMC and SMC, the spatial distribution of the stellar content tends to become more regular and ordered as its age increases. Young stars are found to be results of bursts of star formation, although there is some disagreement on the reasons for the triggered star formation.

Consistent with the previous works are the results from papers based on infrared surveys. Counts of objects towards the Magellanic Clouds from the DENIS near-infrared survey have been studied by \citet{2000A&A...358L...9C}, who used colour-magnitude diagrams (CMD) to distinguish between three groups of objects with different mean ages. The spatial distribution of the three age groups is found to be quite different: the youngest stars exhibit an irregular structure, while the older stars are smoothly and regularly distributed. Moreover, \citet{2000ApJ...542..804N} and \citet{2009A&A...496..375G} have investigated the spatial distribution of the LMC and SMC stellar components, respectively, from 2MASS data. They present morphological analysis of the MCs colour-magnitude diagram, identifying the different populations and estimating their projected spatial distributions of various stellar populations. \citet{2008MNRAS.389..678B} propose a scenario where the present-day, angular-averaged, large-scale structures of both Clouds appear to behave as tidally truncated systems (which is expected, since they are believed to be Milky Way satellites), characterized by well-defined core and halo substructures, even if the Clouds are not spherical systems. Thus, they have undergone severe tidal perturbation when the last dynamical and hydrodynamical interaction between the Clouds took place (about 200 Myr ago, \citet{2007PASA...24...21B}). The older LMC and SMC star clusters, on the other hand, appear to be distributed as an exponential disk. This distribution is possibly reminiscent of the Clouds' structure prior to the last interaction.

Hopefully, some of the questions regarding the Magellanic system can be answered by ESA's cornerstone mission Gaia. It is primarily an astrometric mission, whose goal is to create the largest and most precise three-dimensional map of our Galaxy by providing unprecedented positional and radial velocity measurements for about one billion stars in our Galaxy and throughout the Local Group \citep{2001A&A...369..339P}.
The detection of stars is expected to be complete to V=20 mag (astrometry + photometry), but the radial velocities will not be measured for stars fainter than V=17.5 mag. The astrometric precision should be 25 $\mu$as at V=15 mag and the distances will be known with a precision better than 10\% for hundreds of millions of stars. Due to the spatial resolution nearby galaxies will be resolved in stars. Thus Gaia data will improve estimation of rotational parallaxes for Local Group galaxies, kinematic separation of stellar populations, galaxy orbits, and cosmological history, all very interesting topics of our nearest galaxies. The galaxies that will benefit most from the Gaia survey will be the Magellanic Clouds, since they are nearest to our Galaxy, where millions of stars are expected to be detected by the Gaia instruments.

The main goal of our project is to obtain the spatial distribution of several stellar components in these galaxies, combining the available data sets of various stellar populations. As the nearest neighbours of our Galaxy, the Magellanic Clouds are the most important targets with a large number of stars observed by Gaia. The results of this investigation will also be used to improve the Gaia Universe Model \citep{GUMoverview}, a set of algorithms used by the data generators of the Gaia simulators to generate simulated data in the framework of Gaia data reduction preparation. Our contribution to the Universe Model consists in implementing a model for the galaxies, resolved in stars, as seen by Gaia.
In this paper we present our first results from modelling the density distribution of the Magellanic Clouds based on various available data sets that are appropriate as described in section 2. In section 3 the methodology we follow is explained, and in section 4 the results of this work are presented. Finally our concluding remarks are gathered in section 5.

\section{The data}
To obtain the spatial distribution we used archive data. We selected all-sky surveys and dedicated catalogues, which are homogeneous, have a good coverage of the galaxies, and are deep enough. We are also interested in the distribution of more specific objects, such as carbon stars, or of extended objects like star clusters and associations.

   \begin{figure*}
   \centering
   \includegraphics[width=6cm]{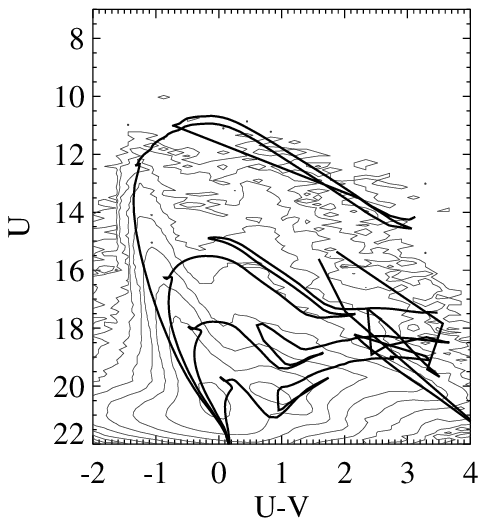}
   \includegraphics[width=6cm]{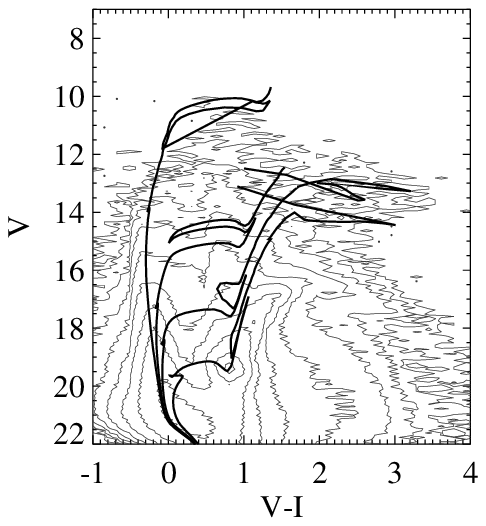}
   \includegraphics[width=6cm]{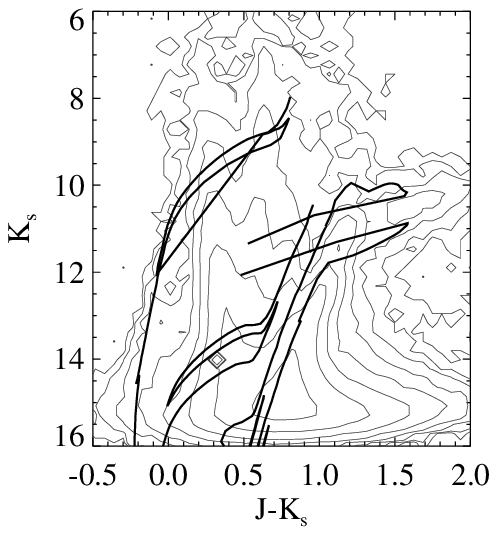}
      \caption{Colour-magnitude diagrams of the LMC from the MCPS (left and middle panel) and 2MASS (right panel). The density levels are logarithmic. Isochrones for 10 Myr, 100 Myr, 300 Myr and 1 Gyr are displayed on top of the contours.}
         \label{fig:cmds}
   \end{figure*}
   
   \begin{table}
      \caption{Age groups}
         \label{tab:age_ranges}
   \centering
   \begin{tabular}{c r c c}
   \hline\hline
     &           Age &       magnitude &                colour \\
   \hline
   A &   $<$ 0.1 Gyr & 11 $<$ U $<$ 16 & -1.5 $<$ U-V $<$ -0.6 \\
   B & 0.1 - 0.3 Gyr & 16 $<$ U $<$ 18 & -1.3 $<$ U-V $<$ 0.2  \\
   C & 0.3 - 0.9 Gyr & 18 $<$ U $<$ 21 & -0.8 $<$ U-V $<$ 1.2  \\
   D &     $>$ 1 Gyr & 17 $<$ V $<$ 20 &  0.4 $<$ V-I $<$ 1.8  \\
   \hline
   E &   $<$ 0.1 Gyr &        K $<$ 15 & -0.5 $<$ J-K $<$ 0.2  \\
   F &     $>$ 1 Gyr & 13.5$<$K$<$15.5 &  0.2 $<$ J-K $<$ 1.2  \\
   \hline
   \end{tabular}
   \end{table}

\subsection{The Magellanic Clouds Photometric Survey (MCPS)}
The Magellanic Clouds Photometric Survey \citep{2002AJ....123..855Z,2004AJ....128.1606Z} offers very rich and complete catalogues of the Magellanic Clouds, but the area they cover is somewhat limited. The MCPS for the SMC by \citet{2002AJ....123..855Z} and for the LMC by \citet{2004AJ....128.1606Z} are catalogues of the U, B, V, and I stellar photometry of only the central 18 deg$^2$ area of the SMC and of the central 64 deg$^2$ area of the LMC. The incompleteness becomes significant at a magnitude fainter than V $<$ 20. This is the reason we are limiting our study to stars brighter than V=20 \footnote{This is convenient since stars fainter than 20th magnitude will not be detected by Gaia because of its observational limit.}. The data are combined with 2MASS and Deep Near-Infrared Southern Sky Survey (DENIS) catalogues to provide, when available, U through K$_s$ data for the stars.

\subsection{The Two-Micron All Sky Survey (2MASS)}
2MASS has uniformly scanned the entire sky in three near-infrared bands to detect and characterize point sources brighter than about 1 mJy in each band, with a signal-to-noise ratio greater than 10. With 2MASS we can choose as large area around the Magellanic Clouds as is necessary. However, the limiting magnitude is somewhat bright with J$<$15.8, H$<$15.1, and Ks$<$14.3 mag \citep{2006AJ....131.1163S}. The data used here are from the 2MASS All-Sky Point Source Catalog, at IPAC Infrared Science Archive (IRSA), Caltech/JPL (http://irsa.ipac.caltech.edu/applications/Gator).

We used the criteria for the main sequence and red giant stars by \citet{gavras2003} (see Table \ref{tab:age_ranges}). These also agree with the criteria used by \citet{2000ApJ...542..804N}.

\subsection{Carbon stars}
We used catalogues of carbon stars in the LMC and SMC from objective-prism plates, taken with the UK Schmidt Telescope. The LMC catalogue is produced by \citet{2001A&A...369..932K}, and the SMC catalogues are from \citet{1993A&AS...97..603R} and \citet{1995A&AS..113..539M}.

Hess diagrams from the various data sets are produced. A Hess diagram shows the relative density of occurrence of stars at different colour-magnitude positions in the colour-magnitude diagram. The colour-magnitude range from the data sets we use was divided in a grid with different numbers of cells for each data set. The size of the cells should provide a fine grid and at the same time contain enough stars for the statistics. The diagrams are shown in Fig.{}\ref{fig:cmds}.  
We divided the stellar content of both Magellanic Clouds into several age groups, matching features of the CMD with isochrones obtained from http://stev.oapd.inaf.it/cmd based on \citet{2008A&A...482..883M} and \citet{1994A&AS..106..275B}.The isochrones for 10 Myr, 100 Myr, 300 Myr, and 1 Gyr are displayed on top of the contours in the Hess diagrams. The selection criteria we use are summarized in Table \ref{tab:age_ranges}.

\section{Methodology}

To determine approximately the shape and the distribution of the galaxies we used two approaches. \\
1. We performed star counts with a rectangular grid. The size of the grid cell was different for the various stellar populations. It was chosen in such a way as to include a sufficient number of stars and also to provide enough detail for the isodensity contour maps. The contours in these maps trace areas with equal stellar density. They show the overall shape of the galaxies and offer an initial insight into their spatial distribution. \\
2. We also used radial density profiles (RDPs) to obtain the actual distribution of the galaxies. The RDPs correspond to the projected radial number density of objects contained in concentric rings around the LMC and SMC centroids. The underlying assumption for this kind of analysis is that the structures should present an important degree of radial symmetry. This is not always the case for the Magellanic Clouds, but RDPs can still be used as probes of the radial distribution of the various objects averaged over all azimuthal directions, hence of the large-scale structure.
We divided the area around each galaxy in concentric annuli of increasing steps outwards and we counted the number of stars within each annulus. The number of stars per unit area in each annulus gives us the stellar density, corresponding to the mean distance from the centre of the galaxy. The thus derived density profiles for each age and/or object group are compared with existing theoretical models

For the centres of the Magellanic Clouds, we adopt the following coordinates given by the SIMBAD Astronomical Database (http://simbad.u-strasbg.fr/simbad) -- for LMC RA: 05$^h$ 23$^m$, Dec: -69$^d$ 45$^m$ and for the SMC RA: 00$^h$ 52$^m$, Dec: -72$^d$ 48$^m$

A conventional first approach to describing the spatial distribution of various stellar populations in the Magellanic Clouds from the RDPs is to  use an exponential-disk profile,
   \begin{displaymath} 
      f (r) = f_{0D} \times e^{ - r / h_D } , 
   \end{displaymath}
where $h_D$ and $f_{0D}$ are the scale length and the central density of objects, respectively, and $r$ is the distance from the centre. As a second approach we use a King-like profile based on the King law \citep{1962AJ.....67..471K}, which is often used to describe the distribution of globular clusters, but also applies to dwarf spheroidal galaxies,
   \begin{displaymath}
      f(r) = f_{0K} \left( \frac{1}{\sqrt{1 + \frac{r^2}{{r_c}^2}}} - \frac{1}{\sqrt{1 + \frac{{r_t}^2}{{r_c}^2}}} \right)^2 , 
   \end{displaymath}
where $r_t$ and $r_c$ are the core and tidal radii, respectively, and $f_{0K}$ the central density of objects and $r$ is the distance from the centre.

The best-fitting profiles are found by performing the Levenberg-Marqwardt least-squares fit to the considered functions. This fit was performed with IDL (Interactive Data Language, http://www.ittvis.com/) with the use of the specially developed procedure MPFITFUN \citep{2009ASPC..411..251M}.

\section{Results}

   \begin{figure*}
   \centering
   \includegraphics[width=7cm]{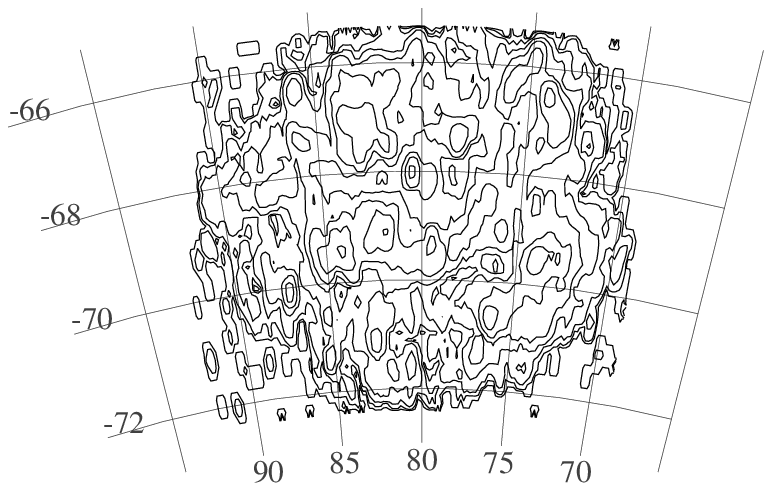}
   \includegraphics[width=7cm]{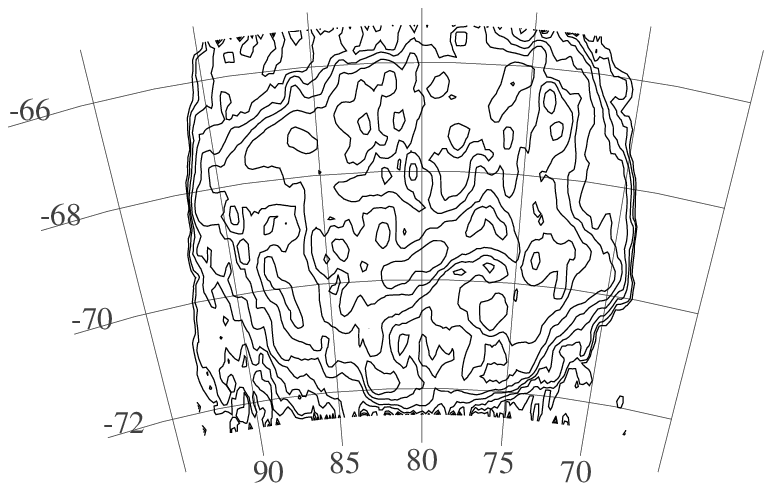}
   \includegraphics[width=7cm]{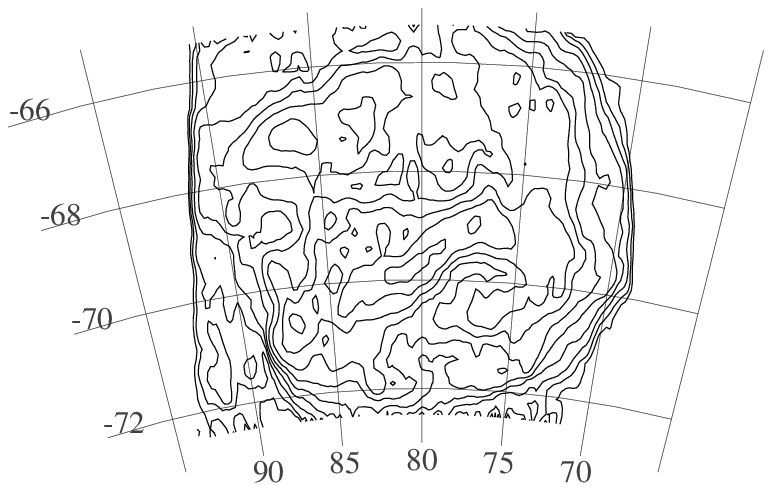}
   \includegraphics[width=7cm]{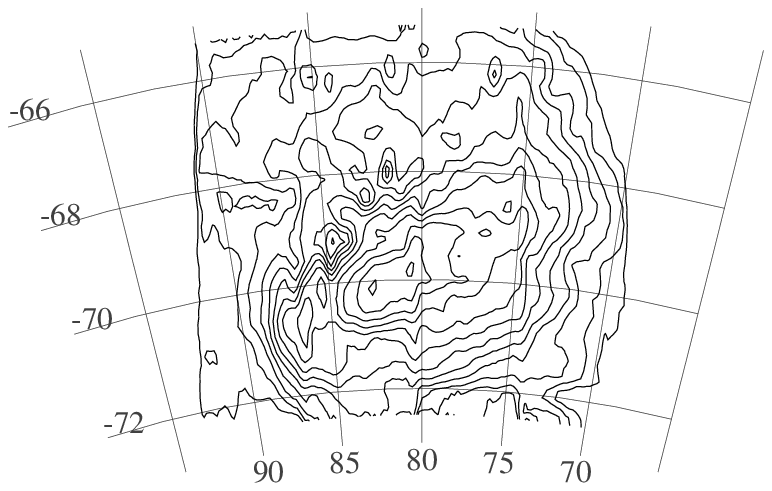}
      \caption{Isodensity contour maps of the LMC stars from the MCPS with various ages. Top row - age groups A and B; bottom row - age groups C and D.}
         \label{fig:iso_opt_lmc}
   \end{figure*}
   
   \begin{figure*}
   \centering
   \includegraphics[width=7cm]{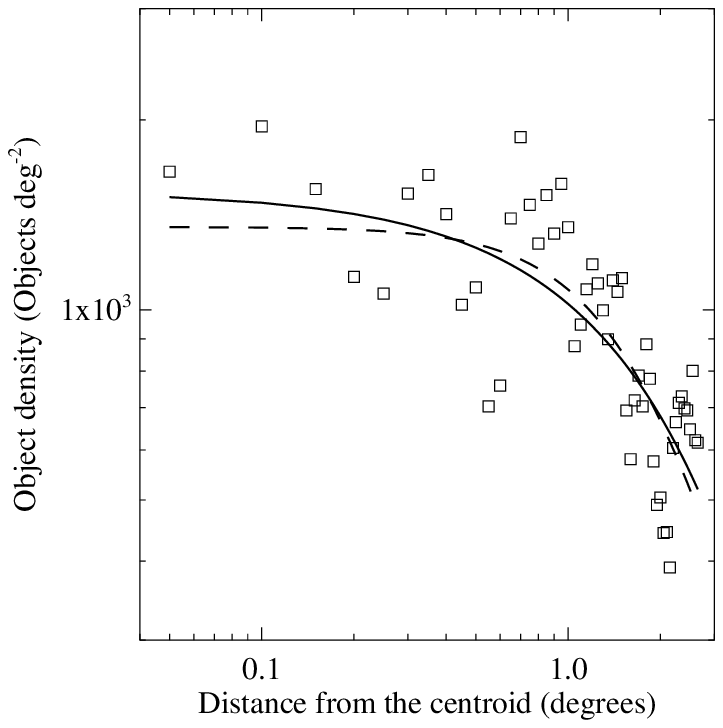}
   \includegraphics[width=7cm]{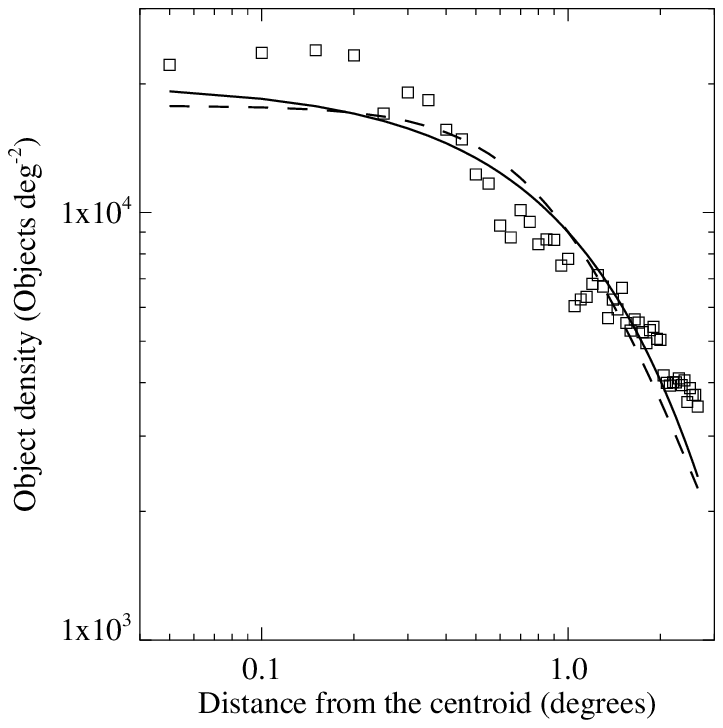}
   \includegraphics[width=7cm]{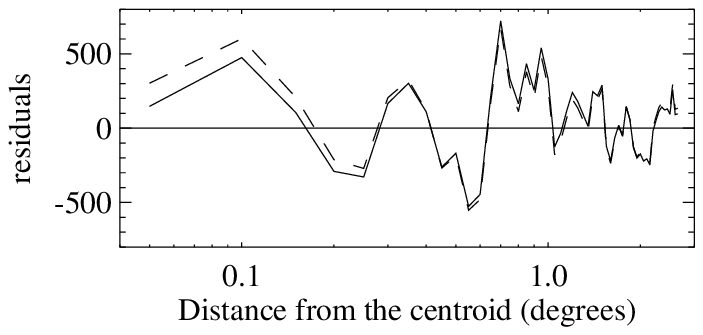}
   \includegraphics[width=7cm]{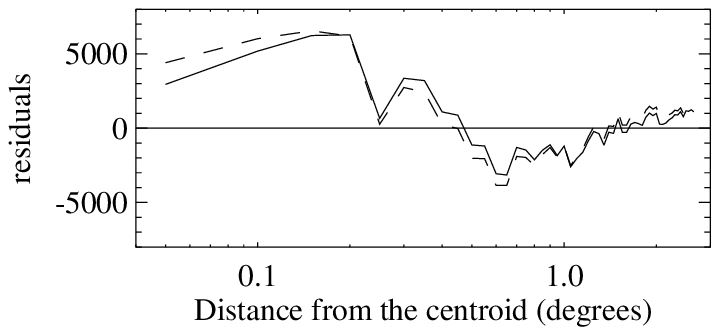}
   \includegraphics[width=7cm]{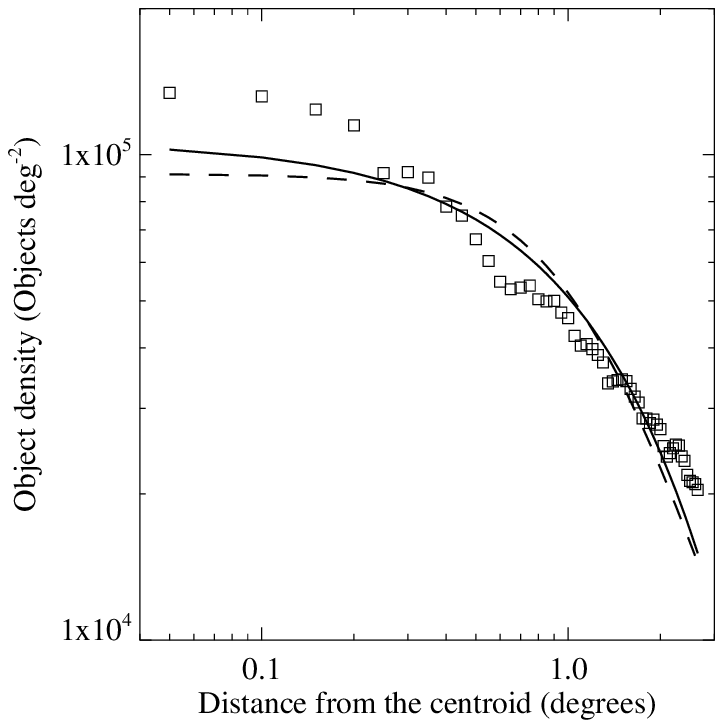}
   \includegraphics[width=7cm]{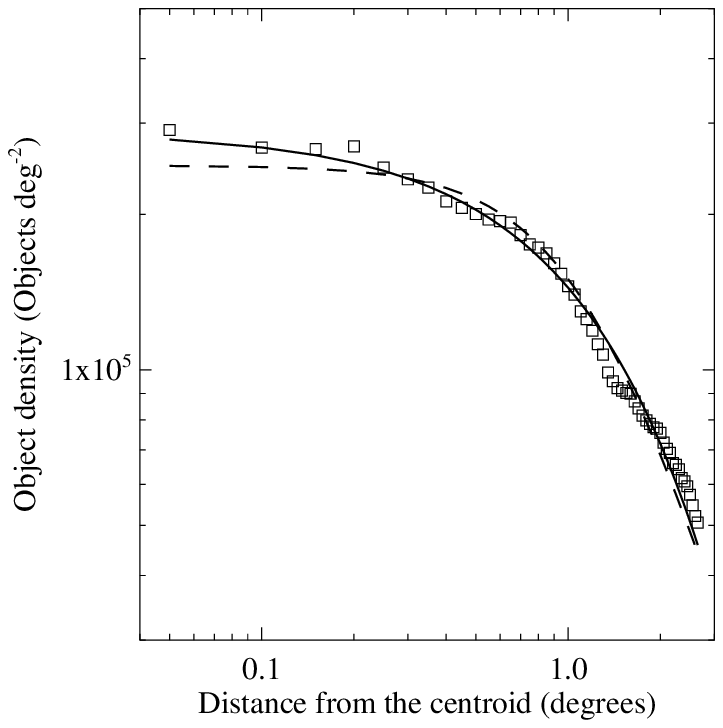}
   \includegraphics[width=7cm]{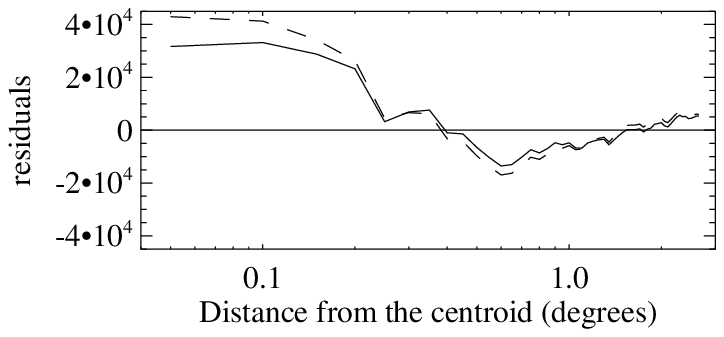}
   \includegraphics[width=7cm]{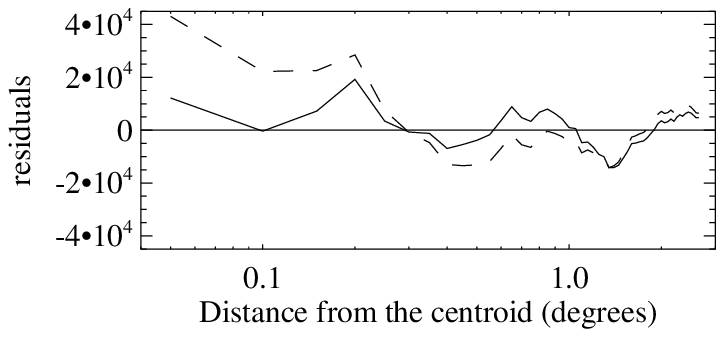}
         \caption{RDPs for the subsets of LMC stars from the MCPS, fitted with exponential-disk (solid line) and King profiles (dashed line). Top row - age groups A and B; bottom row - age groups C and D. Error bars are not shown, because the Poisson errors are comparable in size to or smaller than the symbols. Below each profile are the residuals from the fit.}
         \label{fig:mcps_rdp_lmc}
   \end{figure*}
   
   \begin{figure*}
   \centering
   \includegraphics[width=7cm]{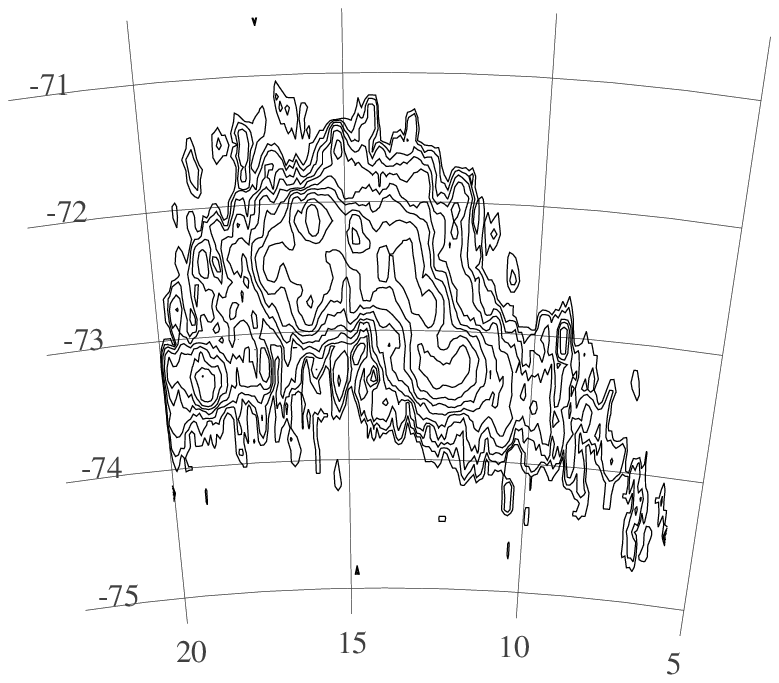}
   \includegraphics[width=7cm]{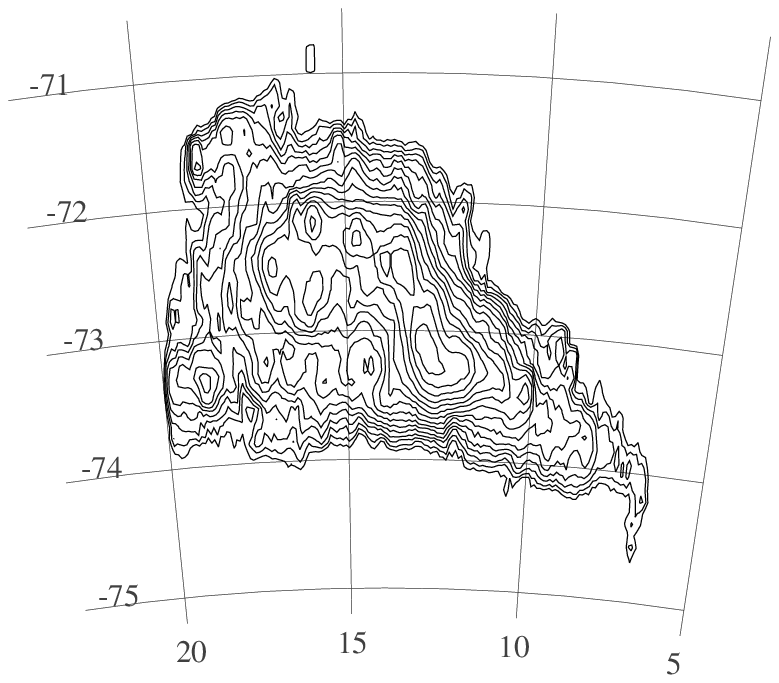}
   \includegraphics[width=7cm]{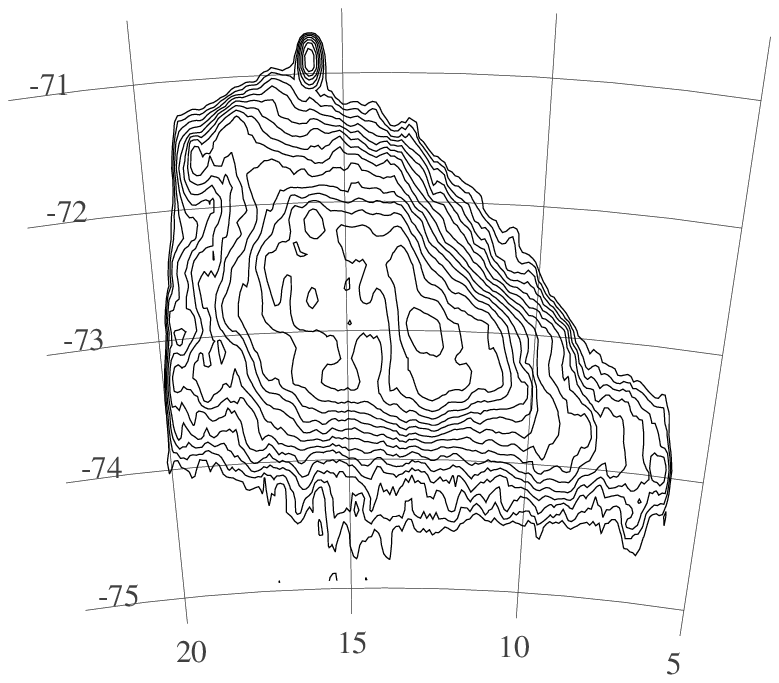}
   \includegraphics[width=7cm]{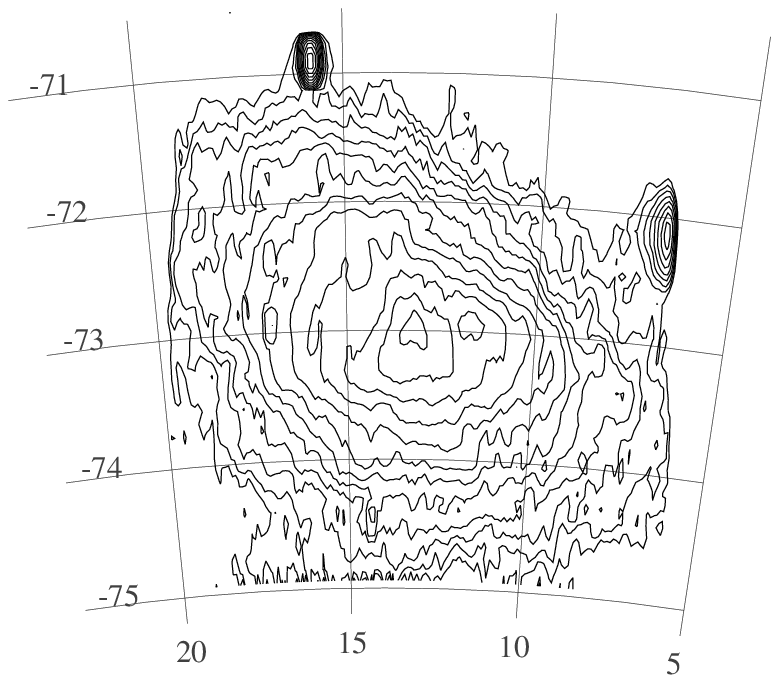}
      \caption{Isodensity contour maps of the SMC stars from the MCPS with various ages. Top row - age groups A and B; bottom row - age groups C and D.}
         \label{fig:iso_opt_smc}
   \end{figure*}

   \begin{figure*}
   \centering
   \includegraphics[width=7cm]{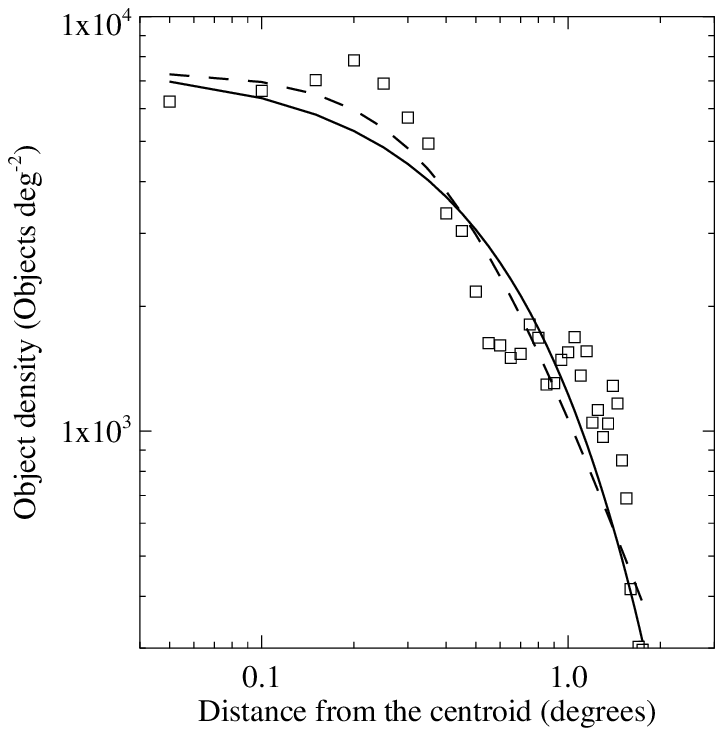}
   \includegraphics[width=7cm]{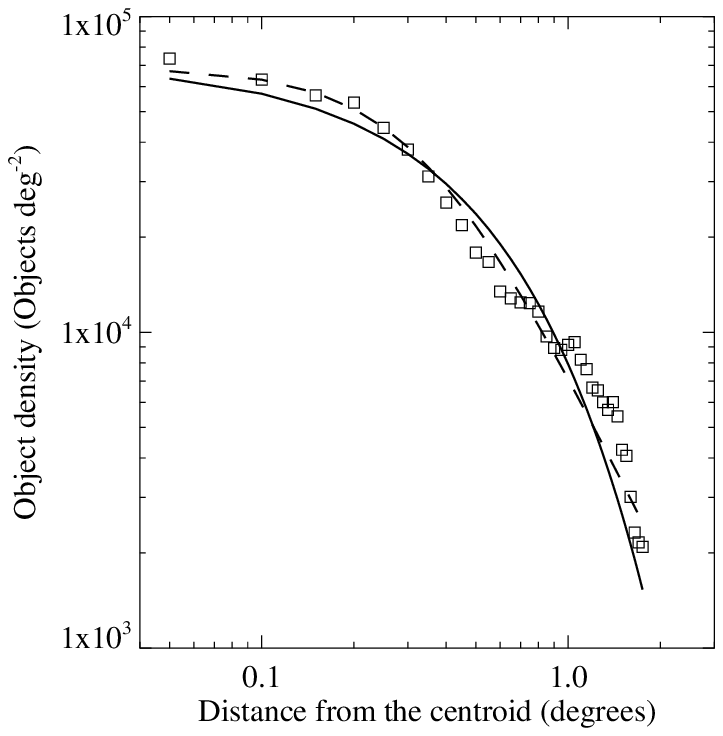}
   \includegraphics[width=7cm]{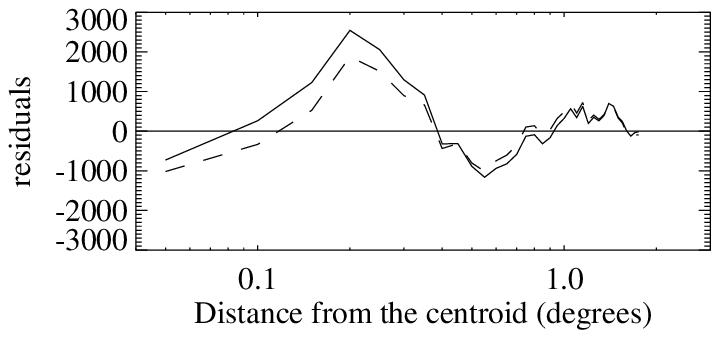}
   \includegraphics[width=7cm]{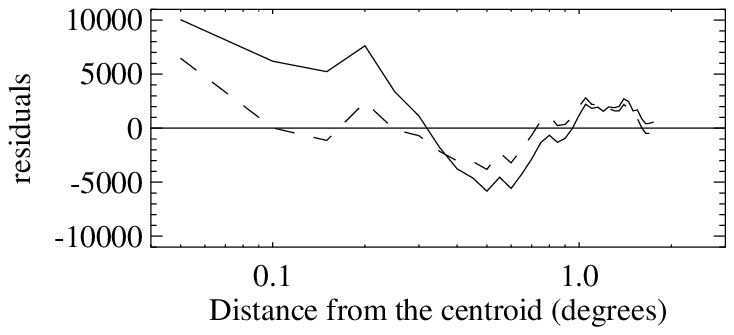}
   \includegraphics[width=7cm]{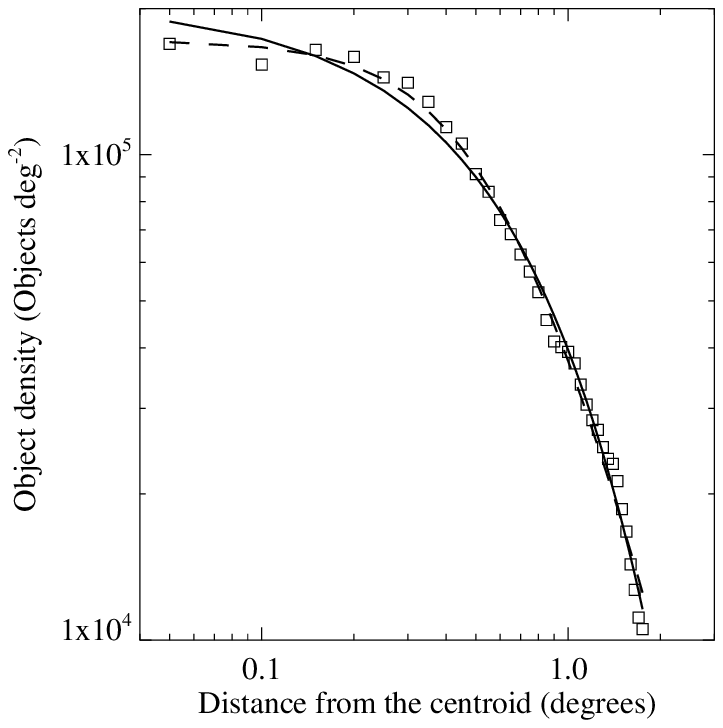}
   \includegraphics[width=7cm]{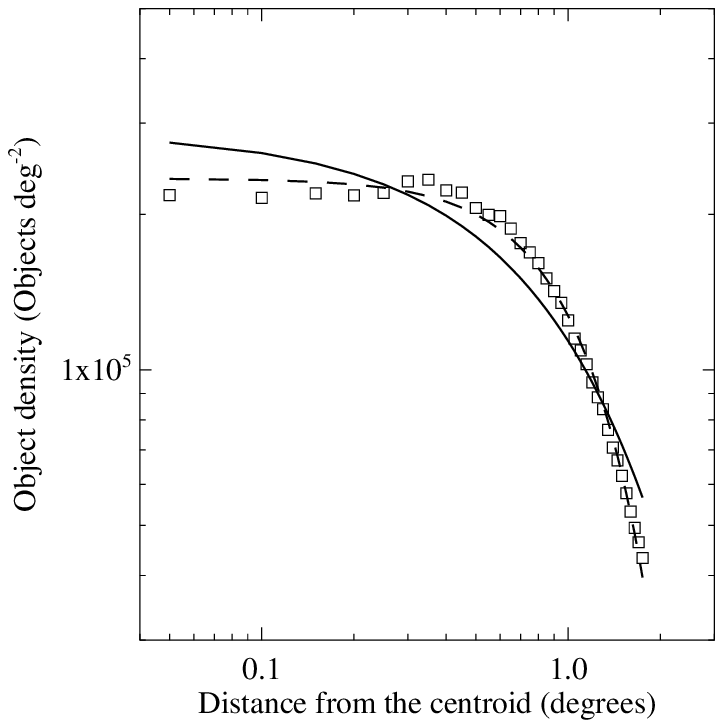}
   \includegraphics[width=7cm]{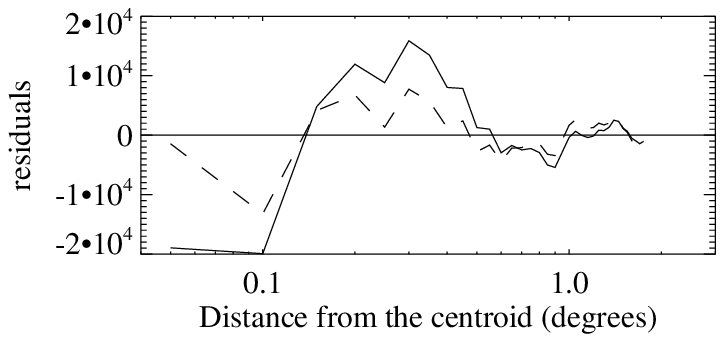}
   \includegraphics[width=7cm]{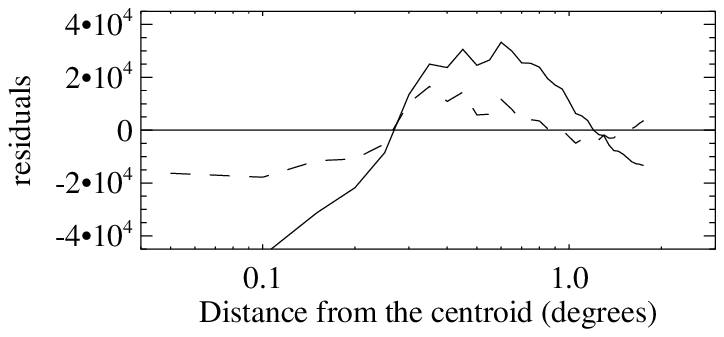}
         \caption{RDPs for the subsets of SMC stars from the MCPS, fitted with exponential-disk (solid line) and King profiles (dashed line). Top row - age groups A and B; bottom row - age groups C and D. Error bars are not shown, because the Poisson errors are comparable in size to or smaller than the symbols. Below each profile are the residuals from the fit.}
         \label{fig:mcps_rdp_smc}
   \end{figure*}

   \begin{figure*}
   \centering
   \includegraphics[width=18cm]{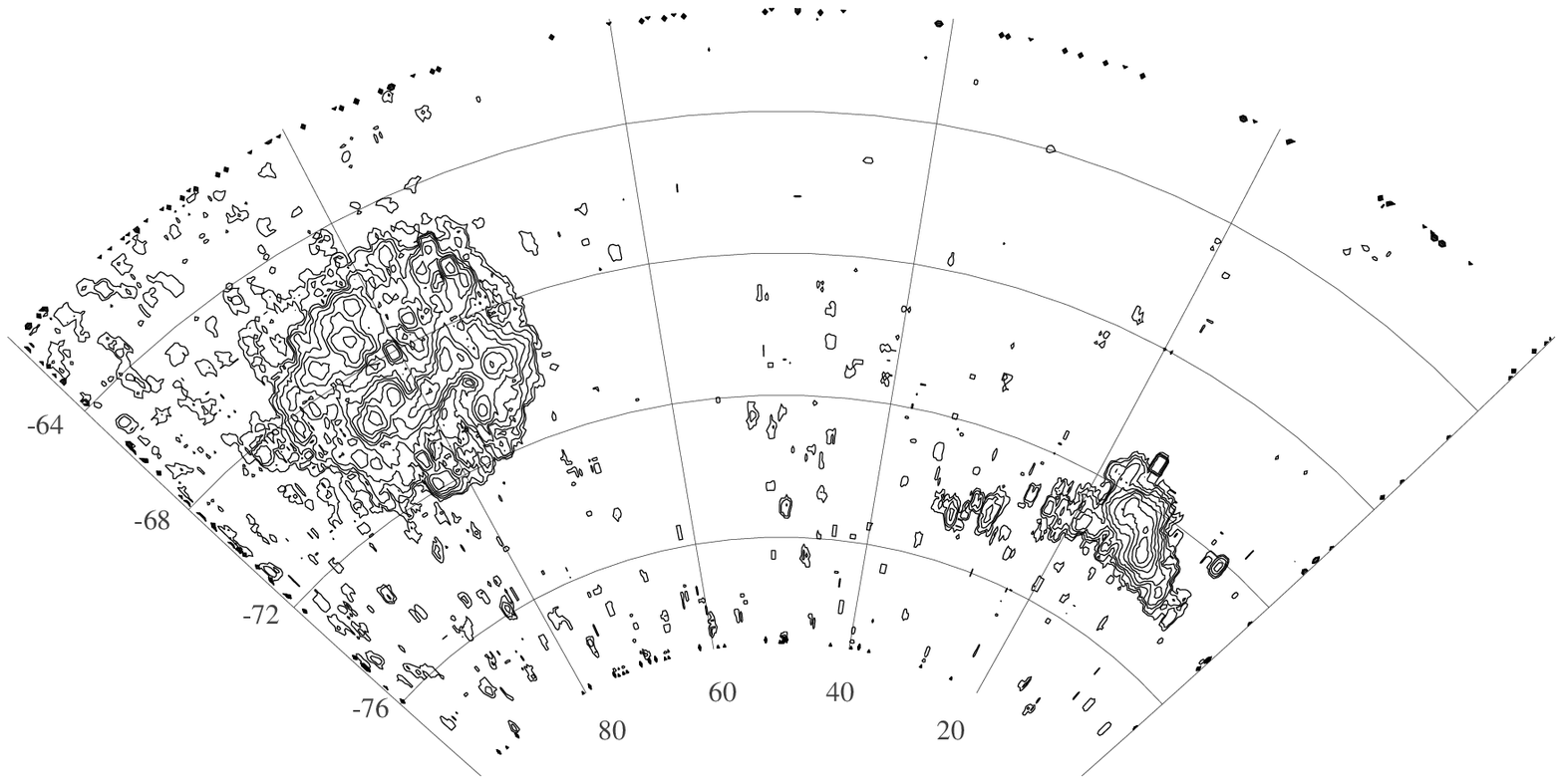} 
   \includegraphics[width=18cm]{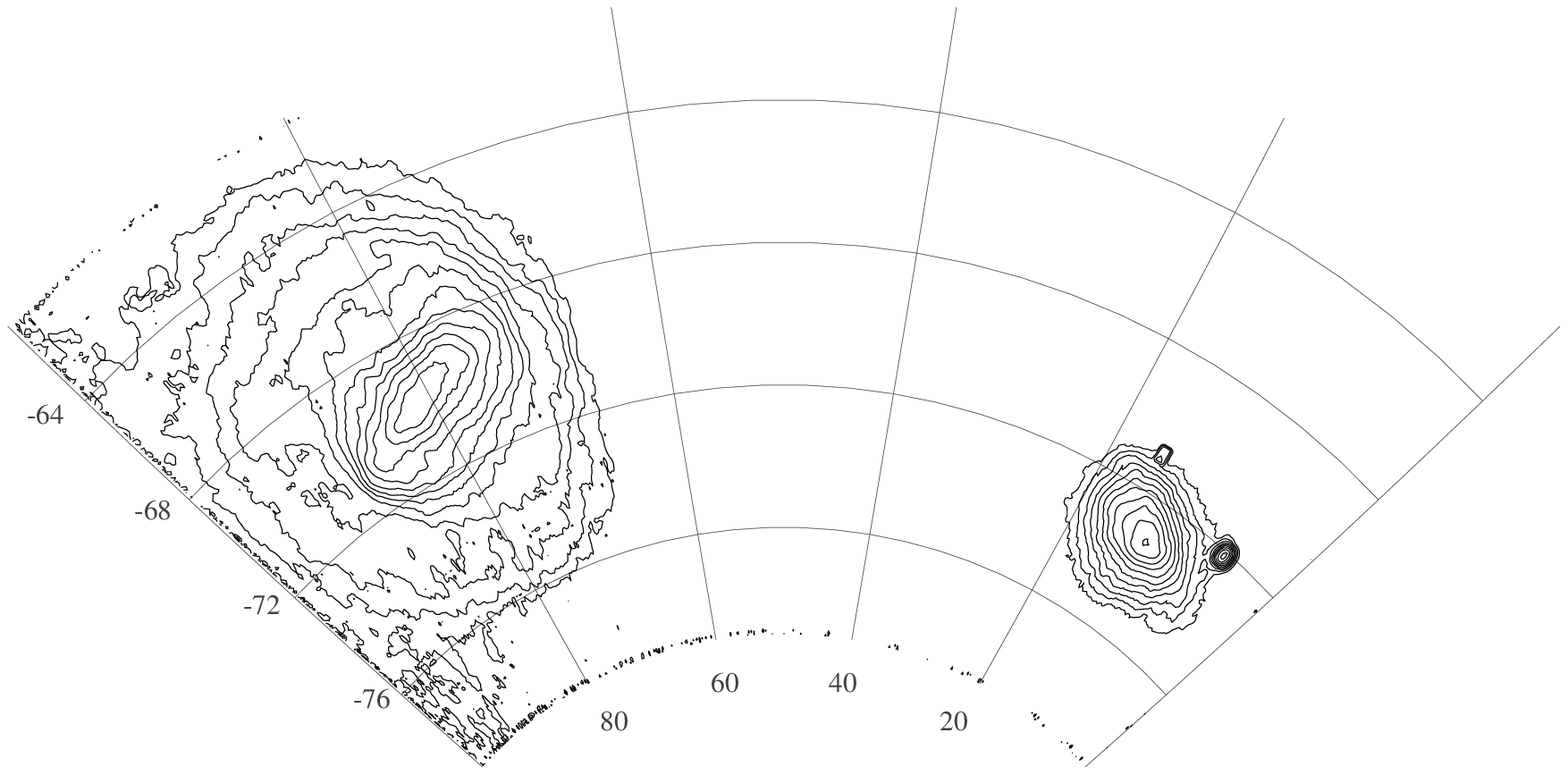} 
      \caption{Isodensity contour maps of the Magellanic Clouds from 2MASS. Top row - age group E; bottom row - age group F. Left - LMC, right - SMC.}
         \label{fig:2mass_iso}
   \end{figure*}
   
   \begin{figure*}
   \centering
   \includegraphics[width=7cm]{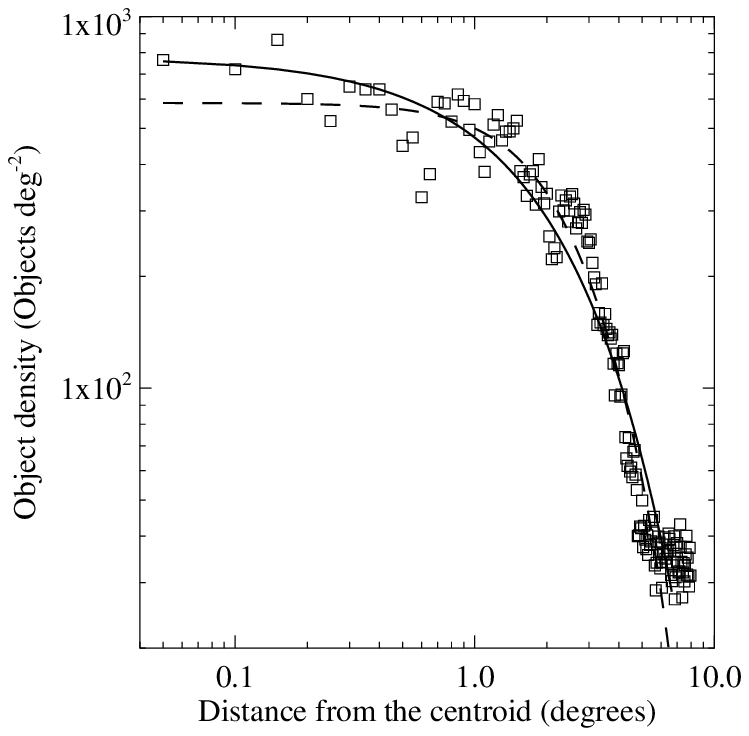} 
   \includegraphics[width=7cm]{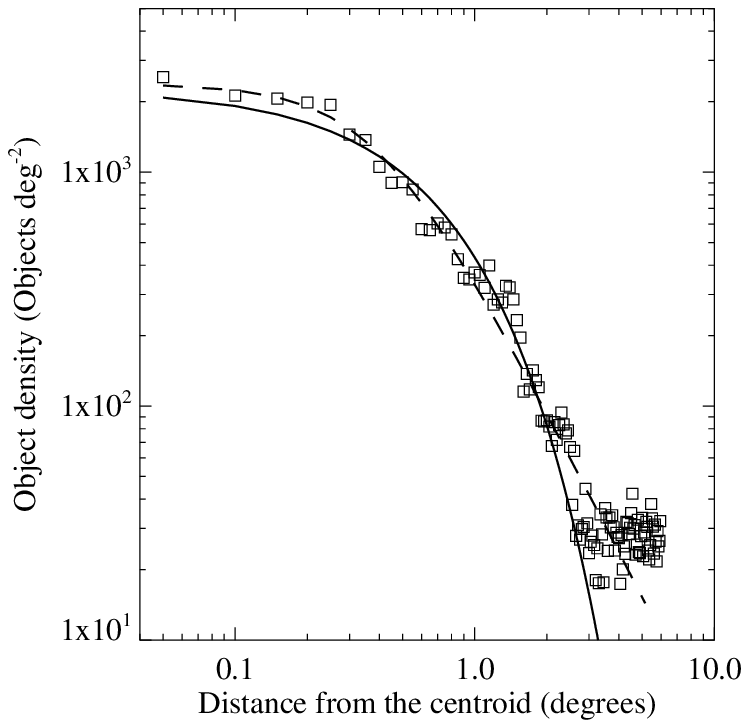}
   \includegraphics[width=7cm]{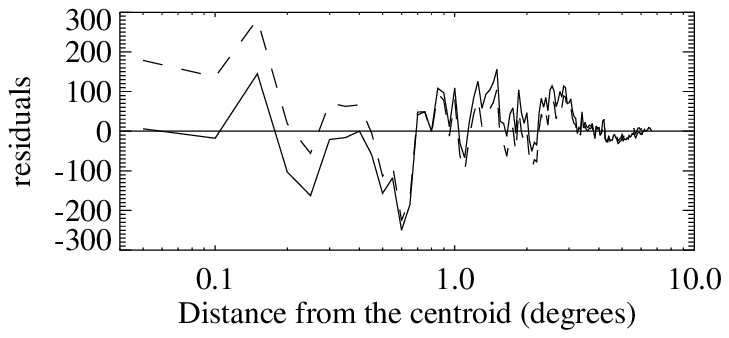}
   \includegraphics[width=7cm]{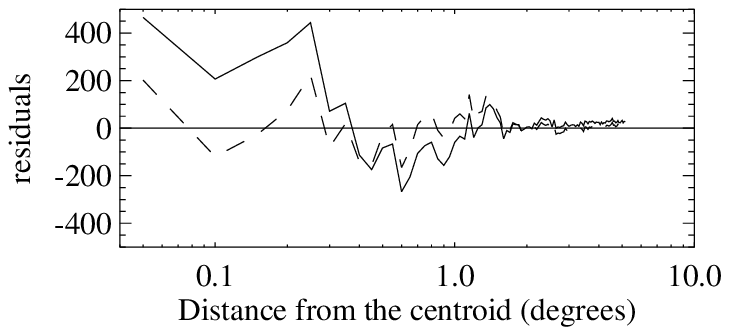}
   \includegraphics[width=7cm]{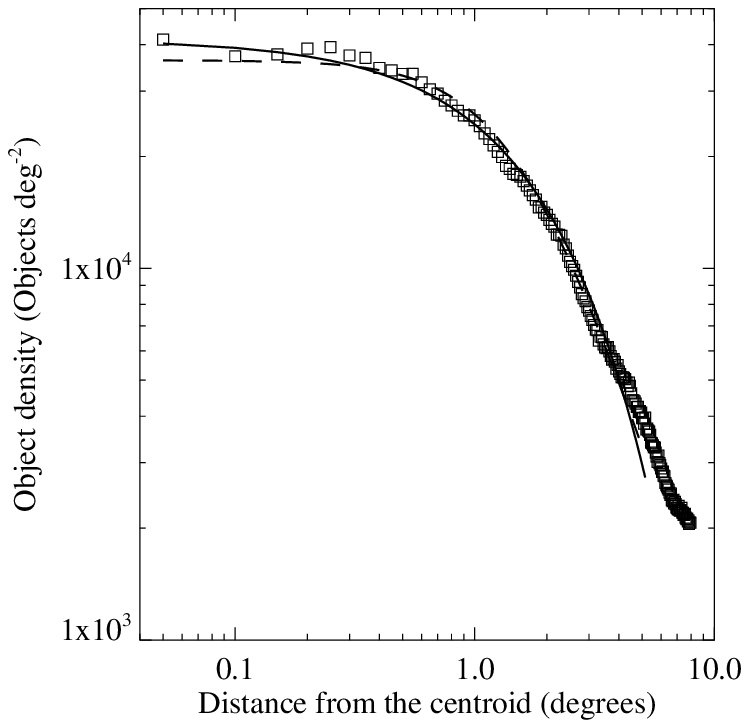} 
   \includegraphics[width=7cm]{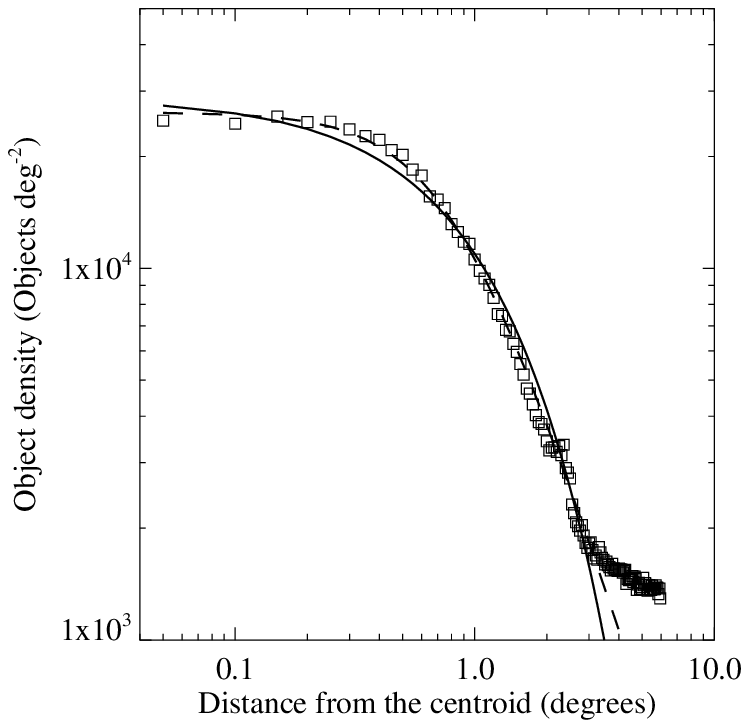}
   \includegraphics[width=7cm]{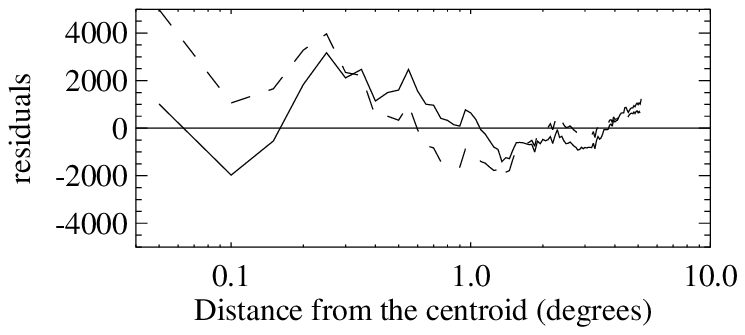}
   \includegraphics[width=7cm]{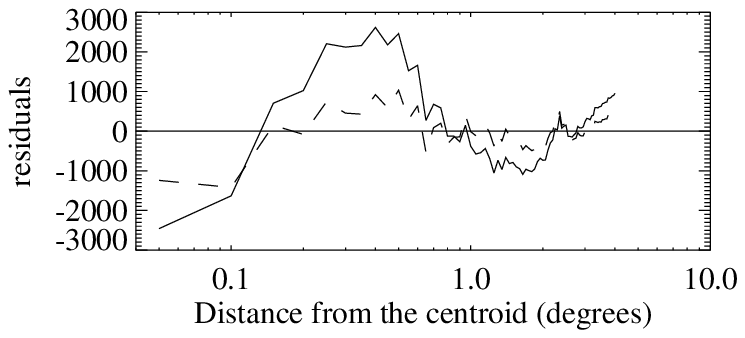}
   \caption{RDPs for the subsets of stars from 2MASS, fitted with exponential-disk (solid line) and King profiles (dashed line). Top row - age group E; bottom row - age group F. Left - LMC; right - SMC. Error bars are not shown, because the poisson errors are comparable in size to or smaller than the symbols. Below each profile are the residuals from the fit.}
         \label{fig:2mass_rdp_all}
   \end{figure*}

   \begin{figure*}
   \centering
   \includegraphics[width=18cm]{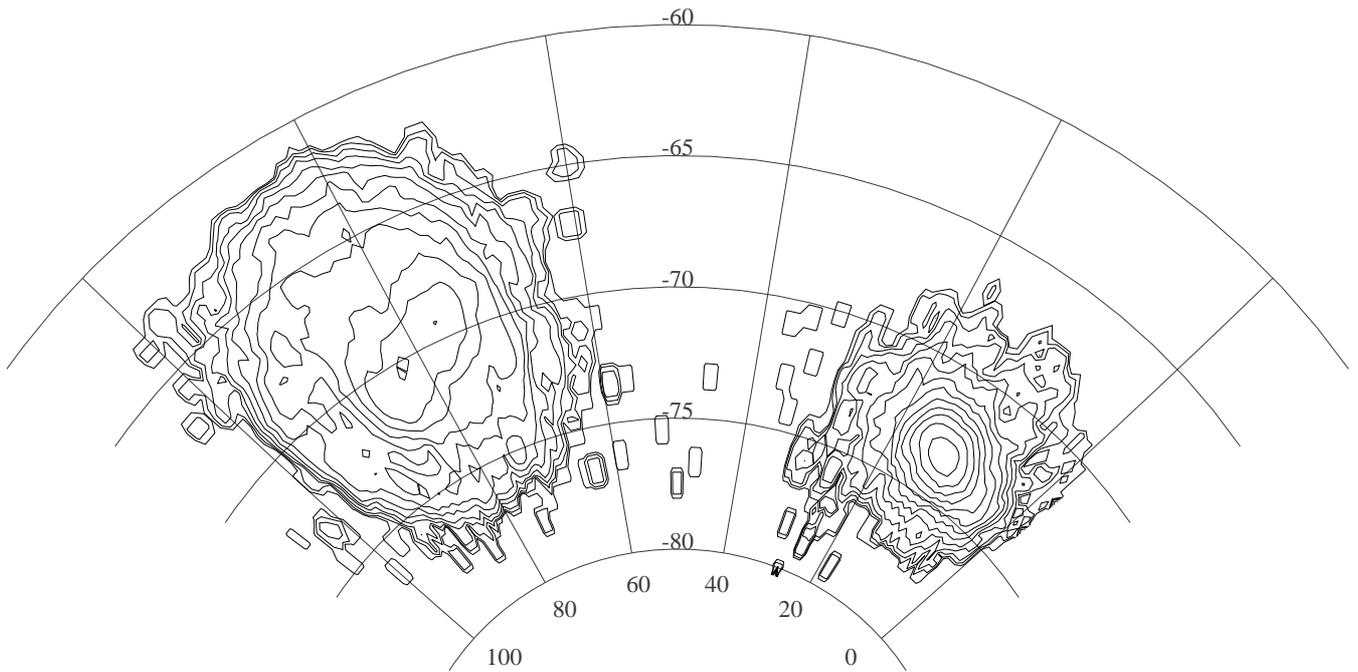} 
      \caption{Isodensity contour maps of the Magellanic Clouds carbon stars. Left - LMC; right - SMC.}
         \label{fig:cs_iso}
   \end{figure*}

   \begin{figure*}
   \centering
   \includegraphics[width=6cm]{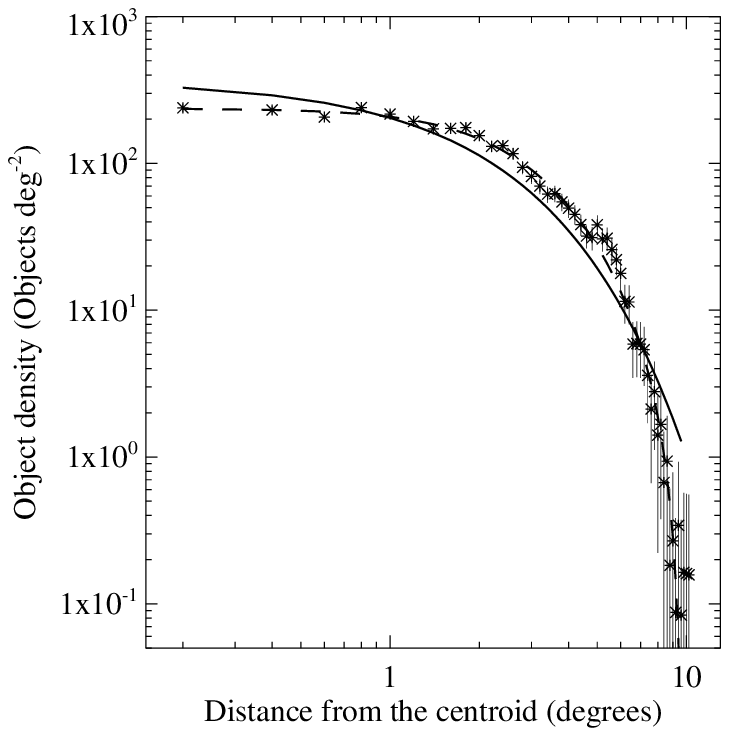}
   \includegraphics[width=6cm]{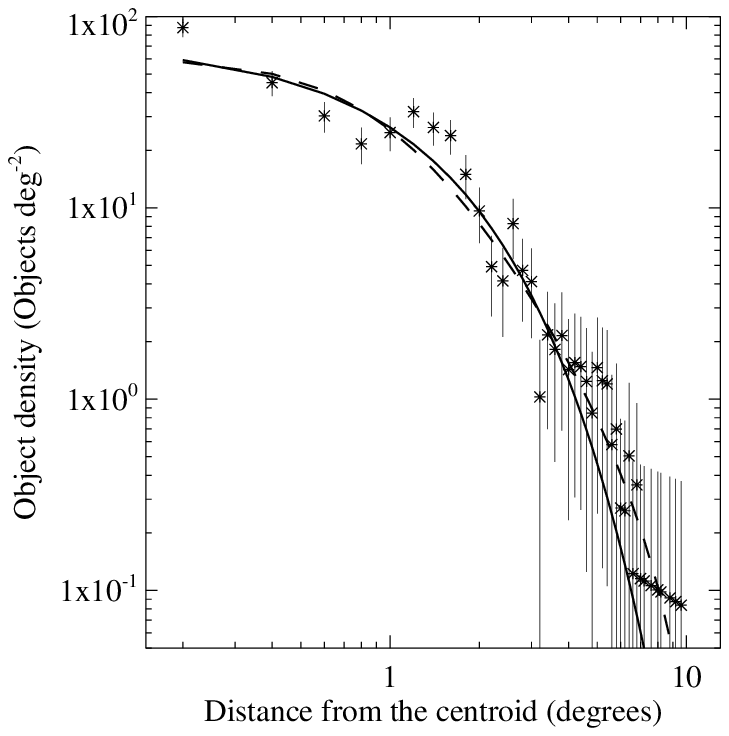}
   \includegraphics[width=6cm]{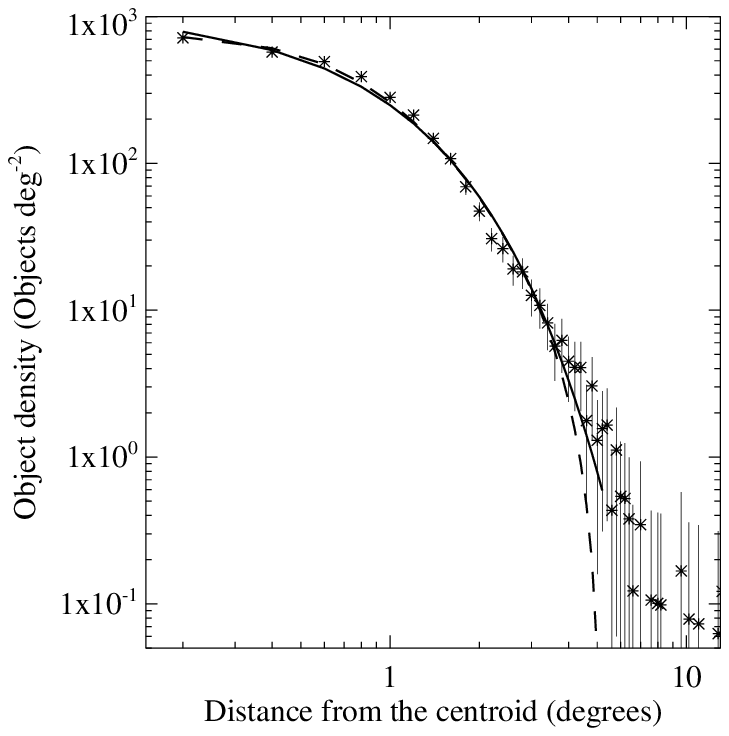}
   \includegraphics[width=6cm]{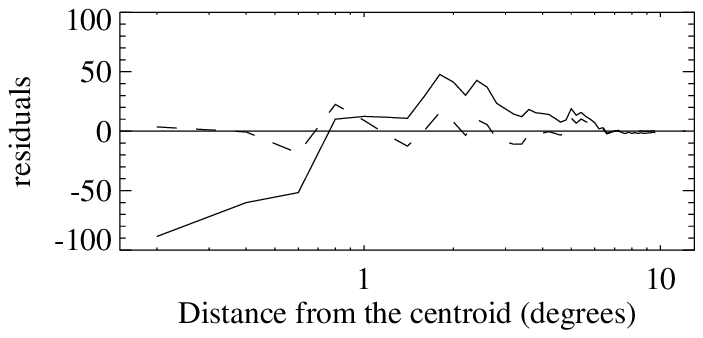}
   \includegraphics[width=6cm]{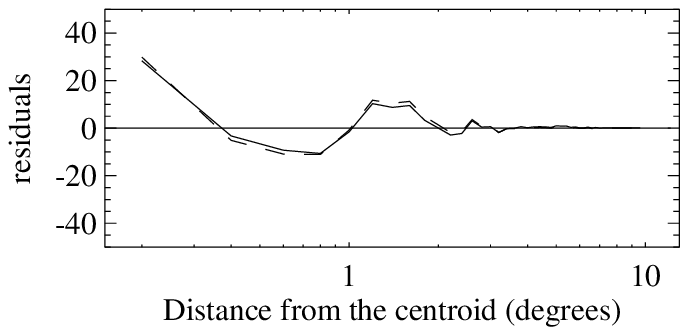}
   \includegraphics[width=6cm]{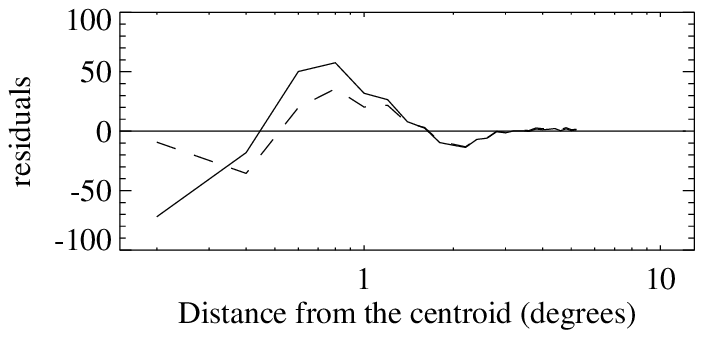}
      \caption{RDPs for the carbon stars, fitted with exponential-disk (solid line) and King profiles (dashed line). Left - LMC; middle - LMC bright; right - SMC. Below each profile are the residuals from the fit.}
         \label{fig:cs_rdp}
   \end{figure*}

Figure \ref{fig:iso_opt_lmc}, \ref{fig:iso_opt_smc}, \ref{fig:2mass_iso}, and \ref{fig:cs_iso} show the isodensity contour maps obtained from the MCPS, 2MASS and carbon star catalogues. The spatial distribution of the various age groups is as expected. The older stars have a more regular and smoother appearance, while younger stars form fragmented and less symmetric structures \citep{2000A&A...358L...9C, 2000ApJ...534L..53Z, 1998A&A...338L..29M, 2001A&A...379..864M, 2009A&A...496..375G}.

The RDPs obtained from MCPS, 2MASS, and the carbon stars are presented in Figure \ref{fig:mcps_rdp_lmc}, \ref{fig:mcps_rdp_smc}, \ref{fig:2mass_rdp_all}, and \ref{fig:cs_rdp}. The best fit of both models is displayed in each figure. In all figures, except Figure \ref{fig:cs_rdp}, error bars are not shown, because the Poisson errors are comparable in size to or smaller than the symbols.

The structural parameters of the LMC and SMC obtained by fitting RDPs from the adopted data sets in the various catalogues are listed in Table \ref{tab:parameters}. Columns (2) and (3) contain the central density of objects $f_{0D}$ and the scale length $h_D$ respectively. Columns (4)-(7) contain the central density of objects $f_{0K}$, the core radius $r_c$, the tidal radius $r_t$ where available, and the concentration parameter $c_p$.
The parametric value of the tidal radius $r_t$ obtained for stars from 2MASS and the MCPS was not reliable enough, because of the relative incompleteness of 2MASS and the small area that is available from the MCPS. The carbon stars, on the other hand, represent a sufficiently complete sample of this particular group of objects as numbers, spatial distribution, and magnitude range. Therefore we can use these objects as prototypes of the major dynamical system in both Clouds.

Figures \ref{fig:mcps_rdp_lmc}a and \ref{fig:mcps_rdp_smc}a display a very clumpy distribution for the brightest stellar population. Considering that the stars in this age group are in concentrations, i.e. stellar complexes \citep{1998A&A...338L..29M,2001A&A...379..864M}, then the very wide spread is expected and explained.

The RDPs for the SMC from MCPS show that both the King and the exponential-disk models can be used to describe the stellar distribution. However, as can be seen from Fig. \ref{fig:mcps_rdp_smc}, if the distribution of the youngest stars is almost equally well fitted by both, the oldest stars prefer the King law. Such a difference is not observed in the LMC, where the exponential-disk model seems slightly better. This is possibly due to the small area available, which is dominated by the younger stars.
A similar behaviour is observed in the RDPs from 2MASS. Both the younger and the older stars in the LMC are distributed on an exponential disk, while the radial density distribution of the SMC is better described by the King profile.

Additionally, the residuals below each profile reveal the preferred model and show the dominant model for each part of the galaxy.

The spatial distribution of stars from 2MASS and of the carbon stars in the LMC (see Figs. \ref{fig:2mass_iso} and \ref{fig:cs_iso}) suggests that there are two subsystems, whose major axes are almost perpendicular to one another. The same behaviour has recently been detected by \citet{2008MNRAS.389..678B} for the star clusters in the LMC. This effect has previously been described by \citet{1990A&AS...84..527K} and later by \citet{1996ApJ...461..742D}, who found that the young clusters, mainly occupying the bar, are rotated with respect to the older clusters.

\begin{table*}
\caption{Structural parameters measured from the RDPs with the exponential-disk and King profiles}
\label{tab:parameters}
\centering
\begin{tabular}{l l c c c c c c}
\hline\hline
  RDP    & $f_{0D}$                      & $h_D $          & $f_{0K}$                      & $r_c$      & $r_t$  & $c_p$\\
         &[Obj.~deg$^{-2}$]              & [deg]           & [Obj.~deg$^{-2}$]             & [deg]      & [deg]  &      \\
  \hline
  \textbf{MCPS} & & & & & & \\  
  LMC A & $(1.4 \pm 0.9) \times 10^{3}$ & $2.6 \pm 0.3$ & $(1.2 \pm 0.6) \times 10^{3}$ & $2.1 \pm 0.2$ & -- & -- \\
  LMC B & $( 18 \pm 1) \times 10^{3}$ & $1.33 \pm 0.07$ & $( 16 \pm 1) \times 10^{3}$ & $ 1.1 \pm 0.06$ & -- & -- \\
  LMC C & $( 96 \pm 4) \times 10^{3}$ & $1.44 \pm 0.06$ & $( 83 \pm 4) \times 10^{3}$ & $ 1.2 \pm 0.07$ & -- & -- \\
  LMC D & $(260 \pm 3) \times 10^{3}$ & $1.52 \pm 0.02$ & $(225 \pm 4) \times 10^{3}$ & $ 1.3 \pm 0.03$ & -- & -- \\
  \hline
  SMC A & $( 7 \pm 0.6) \times 10^{3}$ & $0.57 \pm 0.04$ & $(6.7 \pm 0.5) \times 10^{3}$ & $ 0.4 \pm 0.03$ & -- & -- \\
  SMC B & $(  64 \pm 3) \times 10^{3}$ & $0.48 \pm 0.02$ & $( 62 \pm   2) \times 10^{3}$ & $ 0.4 \pm 0.01$ & -- & -- \\
  SMC C & $( 183 \pm 4) \times 10^{3}$ & $0.64 \pm 0.01$ & $(177 \pm   6) \times 10^{3}$ & $ 0.6 \pm 0.02$ & -- & -- \\
  SMC D & $(259 \pm 11) \times 10^{3}$ & $1.13 \pm 0.06$ & $(918 \pm 247) \times 10^{3}$ & $ 1.9 \pm 0.2 $ & -- & -- \\
  \hline

  \textbf{2MASS} & & & & & & \\  
  LMC E & $( 7.7 \pm 0.2) \times 10^{2}$ & $2.01 \pm 0.05$ & $(11.1 \pm 1.0) \times 10^{2}$ & $2.89 \pm 0.2$& -- & -- \\
  LMC F & $(41.3 \pm 0.5) \times 10^{3}$ & $1.89 \pm 0.02$ & $(36.3 \pm 0.4) \times 10^{3}$ & $1.58 \pm 0.02$& -- & -- \\
  SMC E & $(22.6 \pm 1.0) \times 10^{2}$ & $0.60 \pm 0.01$ & $(23.8 \pm 0.7) \times 10^{2}$ & $0.40 \pm 0.01$& -- & -- \\
  SMC F & $(28.7 \pm 0.7) \times 10^{3}$ & $1.03 \pm 0.02$ & $(26.3 \pm 0.3) \times 10^{3}$ & $0.82 \pm 0.01$& -- & -- \\
  \hline

  \textbf{Carbon stars} & & & & & & \\  
  LMC        &  $368 \pm 22$ & $1.69 \pm 0.07$ &  $514 \pm 17$ &  $3.3 \pm 0.1$  & $9.7 \pm 0.2$ & $0.46$\\
  LMC bright &   $72 \pm 6$  & $0.98 \pm 0.05$ &   $71 \pm 7$  &  $0.9 \pm 0.1$  &  $12 \pm 3$   & $1.12$\\
  SMC        & $1050 \pm 40$ & $0.69 \pm 0.02$ & $1100 \pm 39$ & $0.85 \pm 0.04$ & $5.2 \pm 0.3$ & $0.78$\\
\hline
\end{tabular}
\end{table*}


\section{Discussion \& conclusions}
The structural parameters of the LMC and SMC from the adopted data sets in the various catalogues (MCPS, 2MASS, and carbon stars) are listed in Table \ref{tab:parameters}. Neighter the exponential-disk nor King models are always very explicitly determined as the dominant dynamical model. However, we can adopt a criterion for the acceptance of one or the other model. The values given by the $\chi^2$ test allow us to assume one of the models as more appropriate for describing the distribution of the data. A second criterion could be the age of the stellar population, assuming that the very young population might be distributed on an exponential disk-like structure, even if it does not fit the whole galaxy.

It was shown that there is a system of young objects more concentrated in the central region of the LMC with a position angle almost perpendicular to the other system. Of course the inclination of the LMC is small (almost face-on), allowing the difference to be seen. On the other hand, the SMC has a high inclination to offer clear evidence of such a characteristic. The two systems in the LMC have been previously mentioned, and are explained as one result of the interaction between the Milky Way galaxy and the Magellanic Clouds.

From the carbon stars in the LMC, it was shown that two different systems exist with a core radius of $3.3 \pm 0.1$ deg for the faint and $0.9 \pm 0.1$ deg for the bright carbon stars. An exponential model also shows such a difference in the scale height, $1.69 \pm 0.07$ deg and $0.98 \pm 0.05$ deg, respectively, revealing a smaller central system of more massive carbon stars. The young clusters are also found in a smaller central system, than for the old ones \citep{2008MNRAS.389..678B}. We did not observe such a segregation for the SMC carbon stars.

Although the Magellanic Clouds are assumed to be irregular galaxies, it also seems that the older populations appear to behave as tidally truncated systems, even if they do not show obvious radial symmetry. From Table \ref{tab:parameters} we can see that the parameters of the fitting for the young stellar populations are inconclusive. However, this is no surprise, because even if the bulk of the stars show some radial symmetry, this is not the case for the young bright stars. Both their contour maps and their radial density profiles do not support such an assumption. This is anticipated since clumping of star-forming regions dominates their distribution.

The results of this investigation were used to provide data and model parameters for Gaia simulations and in particular for the Gaia Universe Model.

\begin{acknowledgements}
M.~Belcheva acknowledges financial support from EC FP6 RTN ELSA (MRTN-CT-2006-033481). E.~Livanou would like to thank the State Scholarships Foundation (I.~K.~Y.~) for financial support. G.~Nikolov acknowledges the support of the Bulgarian National Science Research Fund through grant DO 02-85/2008 and partially DO 02-362/2008.
\end{acknowledgements}

\end{document}